\documentclass[12pt,cite]{iopart}
\usepackage{iopams}  
\usepackage{graphicx,epsf,color}

\usepackage{graphicx}
\usepackage{amssymb}

\begin{document}

\title[One-Dimensional Random Walk in Multi-Zone Environment]
{One-Dimensional Random Walk in Multi-Zone Environment}

\author{AV Nazarenko$^1$ and V Blavatska$^{2}$}

\address{$^1$ Bogolyubov Institute for Theoretical Physics  of the National Academy of Sciences of Ukraine, UA--03680 Kyiv, Ukraine }
\address{$^2$ Institute for Condensed Matter Physics of the National
Academy of Sciences of Ukraine, UA--79011 Lviv, Ukraine}
\ead{nazarenko@bitp.kiev.ua}
%\ead{blavatska@itp.uni-leipzig.de}

\begin{abstract}
We study a symmetric random walk (RW) in one spatial dimension in environment, formed by several zones
of finite width, where the probability of transition between two neighboring points and corresponding
diffusion coefficient are considered to be differently fixed. We derive analytically the probability to
find a walker at the given position and time. The probability distribution function is found and has no
Gaussian form because of properties of adsorption in the bulk of zones and partial reflection at the
separation points. Time dependence of the mean squared displacement of a walker is studied as well and
revealed the transient anomalous behaviour as compared with ordinary RW.

\vspace{2pc}

\end{abstract}

% 05.40.Fb Random walks
% 02.50.Ga Markov processes (Probability theory)

\maketitle

\section{Introduction}

The model of random walk (RW) finds its application in the majority of fields like polymer physics, 
economics, computer sciences~\cite{RWbook} and is generally used as simple mathematical formulation of
diffusion process. The standard unconstrained RW is characterized by constant transition probabilities
which are independent on space and time, being the prototype example of a stochastic Gaussian process.

The RW problem in inhomogeneous environment is of great interest since it is connected with physical
phenomena like transport properties in fractures and porous rocks, diffusion of particles in gels,
colloidal solutions and biological cells  (see, e.g., \cite{Havlin87} for a review). 
A particle performing ordinary diffusion is typically characterized by the mean square displacement
$\langle x^2 \rangle$ obeying a scaling law $\langle x^2 \rangle \sim t$ at time $t$ (or, equivalently,
after $t$ steps of RW trajectory) independently of the space dimension $d$ \cite{Van}. Presence of
inhomogeneities may lead to the anomalous diffusion regime with non-linear power-law dependence on $t$.
Numerous analytical and numerical studies have been carried out to analyze the properties of anomalous
diffusion on fractal-like structures like percolation clusters
\cite{Alexander82,Sahimi83,Rammal83,Argyrakis84,Havlin85,Mastorakos93,Lee00,Klemm02}. Anomalous diffusion
is found in processes like the charged particles transport in amorphous semiconductors \cite{Sher75} or
motion of biochemical species in the crowded environment of biological cells \cite{Hofling13}. A space
dependent diffusivity is often observed  in heterogeneous systems, like in processes of Richardson
diffusion in turbulence \cite{Richardson26} or diffusion of tracers in  bacterial and eukaryotic
cells \cite{Platani02}. The non-ergodicity of heterogeneous diffusion processes with space-dependent
diffusion coefficient is demonstrated analytically in Ref. \cite{Cherstvy13}.

Restricting the volume of space available to the diffusing particles by  confining boundaries causes
remarkable entropic effects~\cite{Liu99}. Numerous studies, dealing with the one-dimensional RW in
environment with ``inhomogeneities'' such as walls~\cite{Lehner,Gupta,Percus}, stack of permeable
barriers~\cite{Tanner,BHMST,NFJH} or finite-sized barriers~\cite{Ascher60,Powels}, revealed the physically
significant effects and the non-trivial behaviour with pronounced deviation away from Gaussian one.
In particular, the time dependent (transient) diffusion coefficients are observed in an inhomogeneous
system with parallel walls of arbitrary permeabilities \cite{Tanner}. The breaking of time scale
invariance is found for Brownian random walks in confined space with absorbing walls \cite{Bearup15}.
These studies also point out the difficulties in analytical description in the problems above. Taking
into account the certain properties of such models, we are aiming to generalize.

In the present paper, we focus our attention on the symmetric RW model with position dependent transition
probability which is equal for the left and right steps and varies within the different intervals (zones)
of coordinate axis. Actually, replacing the probabilities of transition and adhesion within each zone with
their mean values, we receive the step-like dependence in space. In particular, the zones of narrow width
in our model can reproduce the walls. 

Thus, we consider RW in environment, consisting of $N$ zones with constant parameters, and reduce the problem
to solving the differential equation with diffusion coefficient which inherits the established space dependence.
We concentrate our attention on finding the probability distribution function (PDF) by neglecting the local
fluctuations of environment characteristics. Since the adhesion property of zones can vary, it is expected
that a time dependence of the averages should differ from that in the case of uniform medium.

 On the basis of PDF found for
any $N$ we calculate $\langle x^2 \rangle$ as function of finite $t$ and investigate its anomalous behaviour
within the three-zone models.
Averaging PDF over time $t$, we can also estimate a probability to find a walker in a given point of space
and compare it with the performed numerical simulations. We interpret its profile peculiarities from the points
of view of transport theory and mechanics.

The layout of the paper is as follows. In the next Section, we constitute the random walk rules and obtain
the differential equation for PDF. The probability function and its distribution (PDF) are found in Section~3.
After calculating the mean square displacement in Section~4, we end up with giving discussion and outlook.

%%%%%%%%%%%%%%%%%%%%%%%%%%%%%%%%%%%%%%%%%%%%%%%%%%%%%%%%%%%%%%%%%%%%%%%%%%%%%%%%%%%%%%%%%%%%%
\section{Formulation of RW Model in $N$-Zone Environment}

We start with discrete RW model, which can be considered as a Markov process, where
unitary step in time results in either zeroth or unitary step in space ${\mathbb Z}$.
The evolution is determined here by stationary transition probability $T(x_{t+1},x_t)$
defined as
\begin{equation}
T(x,y)=p(y)\delta_{x-y,-1}+q(y)\delta_{x-y,1}+r(y)\delta_{x-y,0},\qquad
\sum\limits_{x\in{\mathbb Z}}T(x,y)=1,
\end{equation}
where constraint $p(x)+q(x)+r(x)=1$ is fulfilled for all $x$; $\delta_{x,y}$ is the Kronecker symbol.

Functions $p(x)$ and $q(x)$ determine probabilities to find a walker at points $x-1$ and $x+1$ if it
was at $x$ in previous time moment, respectively; while $r(x)$ corresponds to probability of adhesion
(adsorption) \cite{Gupta}. In general, $p(x)$, $q(x)$ are regarded as arbitrary non-negative functions
which do not exceed the unit.

To formulate our model, we give the finite sequence $\{a_n\}$ of positive numbers,
\begin{equation}
0=a_0<a_1<\ldots<a_{N-1}<a_N=\infty,
\end{equation}
determining the positions of walls or separation points of environment phases.

For $x<0$, we reproduce the same configuration symmetric with respect to the starting point $x=0$.

Using the finite set $\{p_n\}$ of the given positive constants, we consider RW without preference
between the  moves to the left and right  in heterogeneous environment with the following
properties:
\begin{eqnarray}
&&p(x)=p_n\quad {\rm for}\ \ a_{n-1}<|x|<a_n;\qquad p(\pm a_n)=\frac{1}{2}(p_{n+1}+p_n),\\
&&q(x)=p(x),\qquad r(x)=1-2p(x).
\end{eqnarray}
To preserve a probability meaning of the functions, we require $p_n\leq1/2$.

We would like to note that the numerical simulations demonstrate a sensitivity of the particle distribution
in space to definition of transition probability at $x=\{\pm a_n\}$. It says on important role of interface
effect in multi-zone problem.

Time evolution of the probability distribution function (PDF) $P(x,t)$ in Markov picture with our initial
condition $P(x,0)=\delta_{x,0}$ is given by equation:
\begin{equation}
P(x,t+1)=\sum\limits_{y\in{\mathbb Z}}T(x,y)P(y,t),\qquad
\sum\limits_{x\in{\mathbb Z}}P(x,t)=1.
\end{equation}
Master equation with arbitrary distance $\ell$ and time $\tau$ between successive steps reads
\begin{equation}
P(x,t+\tau)=p(x+\ell)P(x+\ell,t)+q(x-\ell)P(x-\ell,t)+r(x)P(x,t).
\end{equation}

To obtain a differential equation for PDF, $t$ should be relatively large while $\ell$ and
$\tau$ tend to zero simultaneously. Using a Taylor expansion of $P(x,t)$, one gets
\begin{equation}\label{diff1}
\partial_tP(x,t)=\upsilon\partial_x[(p(x)-q(x))P(x,t)]+{\cal D}\partial^2_x[(p(x)+q(x))P(x,t)]
+O(\ell^2),
\end{equation}
here $\upsilon\equiv\ell/\tau$ and ${\cal D}\equiv\ell^2/2\tau$ are constant drift velocity
and diffusivity, respectively.

Omitting terms of order $O(\ell^2)$ and fixing $\upsilon=1$ and ${\cal D}=1/2$,
we obtain differential equation of RW in diffusion approximation:
\begin{equation}\label{idiff}
\partial_tP(x,t)=\partial^2_x[D(x)P(x,t)],\qquad
D(x)=\sum\limits_{n=1}^N p_n\chi_n(x),
\end{equation}
where diffusivity function $D(x)$ coincides with $p(x)$, and
\begin{equation}
\chi_1(x)=\left\{
\begin{array}{cl}
1, & |x|<a_1\\
1/2, & |x|=a_1\\
0, & {\rm otherwise}
\end{array}
\right.,\quad
\chi_{n>1}(x)=\left\{
\begin{array}{cl}
1, & a_{n-1}<|x|<a_n\\
1/2, & |x|=a_{n-1},a_n\\
0, & {\rm otherwise}
\end{array}
\right..
\end{equation}

Characteristic functions $\{\chi_n\}$ of intervals are orthogonal and have the properties:
\begin{eqnarray}
&&\sum\limits_{n=1}^N\chi_n(x)=1,\quad |x|<\infty;\qquad
\chi_n(x)\chi_m(x)=0,\quad n\not=m;\label{cond2}\\
&&\int_{-\infty}^{\infty}\chi_n(x)\chi_m(x)\rmd x=2(a_n-a_{n-1})\delta_{n,m},
\qquad n,m<N.
\end{eqnarray}

However, in the continuous limit functions $\delta_{x,y}$ and $\chi_n(x)$ should be replaced with
distributions, using Dirac $\delta$-function and Heaviside $\theta$-function. We find that
\begin{equation}
P(x,0)=\delta(x),\qquad
D(x)=p_1+\sum\limits_{n=1}^{N-1}\alpha_n\theta(|x|-a_n),
\end{equation}
where $\alpha_n=p_{n+1}-p_n$, $|\alpha_n|<1/2$; and $\theta(x)=[1+\mathrm{sign}(x)]/2$.

Although $D(x)$ contains information on geometry and phases of environment as ``input'',
PDF is evolving in space-time and determines a normalized statistical measure $\mu_t$ for a fixed $t$:
\begin{equation}
\rmd\mu_t=P(x,t)\rmd x,\qquad
\int\rmd\mu_t=1.
\end{equation}

In practice, we compute the probability of finding a walker at point $x$ at time $t$,
\begin{equation}\label{defPr}
{\rm Pr}(x,t)=\frac{1}{t}\int_0^tP(x,\tau)\rmd\tau,
\end{equation}
that is the frequency of visiting a point $x$ in time $t$.

%%%%%%%%%%%%%%%%%%%%%%%%%%%%%%%%%%%%%%%%%%%%%%%%%%%%%%%%%%%%%%%%%%%%%%%%%%%%%%%%%%%%%%%%%%%%%
\section{Finding a Solution}

Developing an approach with $p_n>0$, let us consider the functions
\begin{equation}
D(\alpha,x)=\sum\limits_{n=1}^N (p_n)^{\alpha}\chi_n(x),\qquad
D(0,x)=1,
\end{equation}
which are generated by the diffusivity function $D(x)\equiv D(1,x)$ and depend on both $x$ and $\{p_n\}$
simultaneously, what requires specific rules of differential calculus.

Due to orthogonality of basis $\{\chi_n(x)\}$, one has
\begin{equation}
D(\alpha,x)D(\beta,x)=D(\alpha+\beta,x),\qquad
x\in\mathbb{R}/\{\pm a_n\}.
\end{equation}

Differentiation (integration) of $D(\alpha,x)$ in space $x$ is mainly related with a change of the basis,
while exponents of parameters $\{p_n\}$ stay unchanged. One has
\begin{equation}
D^{\prime}(\alpha+\beta,x)=D^{\prime}(\alpha,x)D(\beta,x)+D(\alpha,x)D^{\prime}(\beta,x),
\end{equation}
where prime symbol means differentiation with respect to coordinate.

Numerical coefficients in the last formula are subjects of relation:
\begin{eqnarray}
(p_{n+1})^{\alpha+\beta}-(p_n)^{\alpha+\beta}&=&
\frac{1}{2}\left[(p_{n+1})^{\alpha}-(p_n)^{\alpha}\right]\left[(p_{n+1})^{\beta}+(p_n)^{\beta}\right]
\nonumber\\
&+&\frac{1}{2}\left[(p_{n+1})^{\alpha}+(p_n)^{\alpha}\right]\left[(p_{n+1})^{\beta}-(p_n)^{\beta}\right],
\label{DD'}
\end{eqnarray}
where $1/2$ and sign ``$+$'' in brackets result from $\theta$-function in $D$, while ``$-$'' results from
$D^\prime$ containing $\delta$-function.

To find PDF, we first concentrate the geometrical data in the new coordinate~\cite{Gungor}
\begin{equation}\label{yfun}
y(x)=\int_0^xD(-1/2,x^{\prime})\rmd x^{\prime}.
\end{equation}
Although there exists one-to-one correspondence between $x$ and $y$, the derivatives of $y(x)$
are singular, in general, at the points $x=\{\pm a_n\}$.

Integrating (\ref{yfun}), we obtain that $y(x)={\rm sign}(x)Y(-1/2,|x|)$, where
\begin{equation}\label{Yfunc}
Y(\alpha,x)\equiv\frac{1}{2}\sum\limits_{n=1}^N(p_n)^{\alpha}\left(a_n-a_{n-1}+|x-a_{n-1}|-|x-a_n|\right),
\quad
x\geq0.
\end{equation}

At this stage, let us define the new functions ${\tilde D}(\alpha,y(x))=D(\alpha,x)$:
\begin{equation}
{\tilde D}(\alpha,y)=\sum\limits_{n=1}^N(p_n)^{\alpha}\tilde\chi_n(y),
\end{equation}
where $\tilde\chi_n(y)=\theta(|y|-b_{n-1})-\theta(|y|-b_n)$ and $b_n\equiv y(a_n)$.

Then, introducing distribution ${\cal P}(y,t)$, we re-write statistical measure  as
\begin{eqnarray}
\rmd\mu_t&=&{\cal P}(y,t)\rmd y
\nonumber\\
&=&{\cal P}(y(x),t){\tilde D}(-1/2,y(x))\rmd x.
\end{eqnarray}

Substituting the re-defined $P(x,t)$ and $\partial_x={\tilde D}(-1/2,y)\partial_y$ into (\ref{idiff}), we have
\begin{equation}\label{DP}
\partial_t{\cal P}-\partial^2_y{\cal P}=\kappa\partial_y(\beta{\cal P}),\qquad
{\cal P}(y,0)=\delta(y),
\end{equation}
where constant $\kappa$ regulates an interface effect between zones.

The right hand side of (\ref{DP}) can be reduced to the form with
$\beta(y)={\tilde D}(-1/2,y)\partial_y{\tilde D}(1/2,y)$. Computations lead to expressions:
\begin{eqnarray}
&\beta(y)=\mathrm{sign}(y)\sum\limits_{n=1}^{N-1}\beta_n\delta(|y|-b_n),\qquad
&\beta_n=\frac{p_{n+1}-p_n}{2\sqrt{p_np_{n+1}}},\label{beta}\\
&B(y)=\sum\limits_{n=1}^{N-1}\beta_n\theta(|y|-b_n),
&\partial_yB(y)=\beta(y).
\end{eqnarray}

Taking into account the specific properties of $D$-functions which should be investigated
in more details, the form of $\beta_n$ can vary what influences also the value of
$\kappa$. However, a sign of $\beta_n$ is defined by difference $p_{n+1}-p_n$. Moreover, $B(y)$
is not logarithmic function here, although $\beta(y)$ is similar to its log-derivative by construction.

We have already found that the bulk diffusion mainly depends on $y(x)$. Remaining problem
is to account for the surface effect determined by $\beta(y)$ and $\kappa$. We attempt to find
${\cal P}(y,t)$ within the perturbation scheme, when
\begin{equation}
{\cal P}(y,t)=\varphi(y,t)+\sum\limits_{r=1}^{\infty}\kappa^r{\cal S}_r(y,t),
\end{equation}
where
\begin{equation}
\varphi(y,t)=\frac{1}{\sqrt{4\pi t}}\exp{\left(-\frac{y^2}{4t}\right)}
\end{equation}
is Gaussian distribution resulting in probability (\ref{defPr}):
\begin{equation}
\Phi(y,t)=\frac{1}{\sqrt{\pi t}}\exp{\left(-\frac{y^2}{4t}\right)}+
\frac{|y|}{2t}\left[{\rm erf}\left(\frac{|y|}{2\sqrt{t}}\right)-1\right].
\end{equation}

A possible solution of  inhomogeneous equation
$\partial_t{\cal S}_1-\partial^2_y{\cal S}_1=\partial_y(\beta\varphi)$
is the function  ${\cal S}_1(y,t)=-B(y)\varphi(y,t)-{\cal F}(y,t)$, where
\begin{equation}
{\cal F}(y,t)=-\frac{1}{2}\sum\limits_{n=1}^{N-1}\beta_n
[\varphi(|y-b_n|+b_n,t)+\varphi(|y+b_n|+b_n,t)].
\end{equation}
Finding ${\cal F}(y,t)$ from the auxiliary equation
$\partial_t{\cal F}-\partial^2_y{\cal F}=\beta\partial_y\varphi$, we contract
$\beta\partial_y\varphi$ and $\varphi$ playing a role of fundamental solution to
diffusion equation with the use of formula:
\begin{equation}\label{contract}
\int_0^t\exp{\left(-\frac{y^2_1}{4(t-\tau)}-\frac{y^2_2}{4\tau}\right)}\frac{\rmd\tau}{\sqrt{(t-\tau)\tau^3}}
=\sqrt{\frac{4\pi}{y^2_2t}}\exp{\left[-\frac{(|y_1|+|y_2|)^2}{4t}\right]}.
\end{equation}

Limiting ourselves by accounting for the first-order correction, we obtain our main result for PDF:
\begin{eqnarray}
&&P(x,t)=\varphi(y(x),t)[1-\kappa B(y(x))]D(-1/2,x)
\nonumber\\
&&+\frac{\kappa}{2}\sum\limits_{n=1}^{N-1}\beta_n
[\varphi(|y(x)-b_n|+b_n,t)+\varphi(|y(x)+b_n|+b_n,t)]D(-1/2,x).
\label{dist}
\end{eqnarray}
This formula looks applicable at $|\beta_n|<1$ because the correction is of the same order
of magnitude as $\varphi$. The PDF of ordinary RW is restored at $2p_n=1$ for all $n$,
when $\beta_n=0$ and $y(x)=\sqrt{2}x$.

\begin{figure}
\begin{center}
\includegraphics[width=7.7cm]{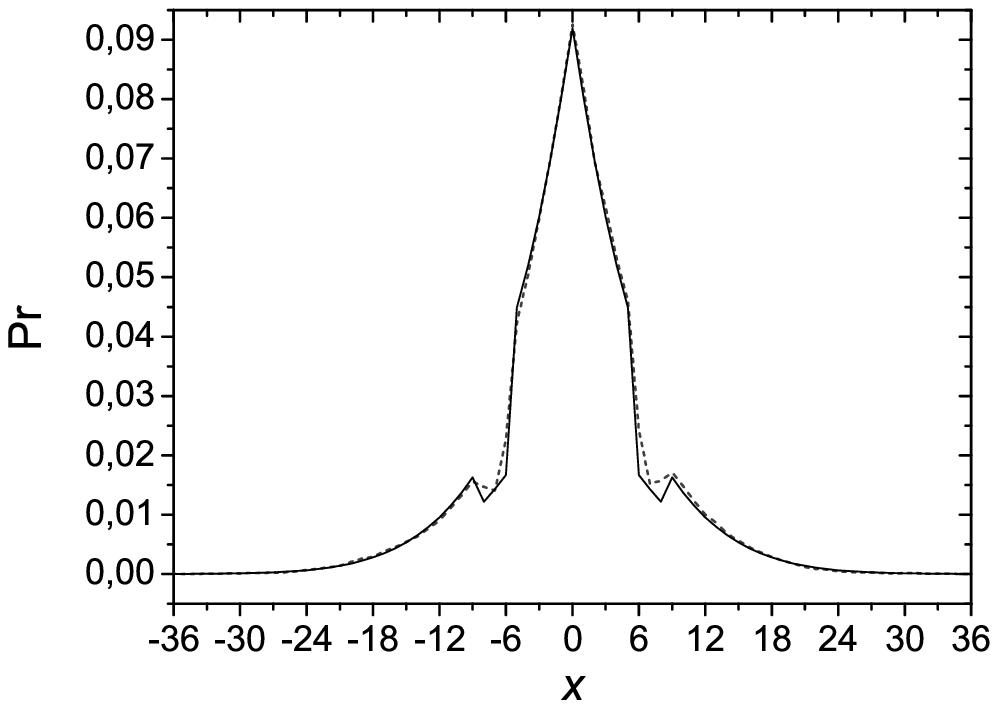}
\includegraphics[width=7.7cm]{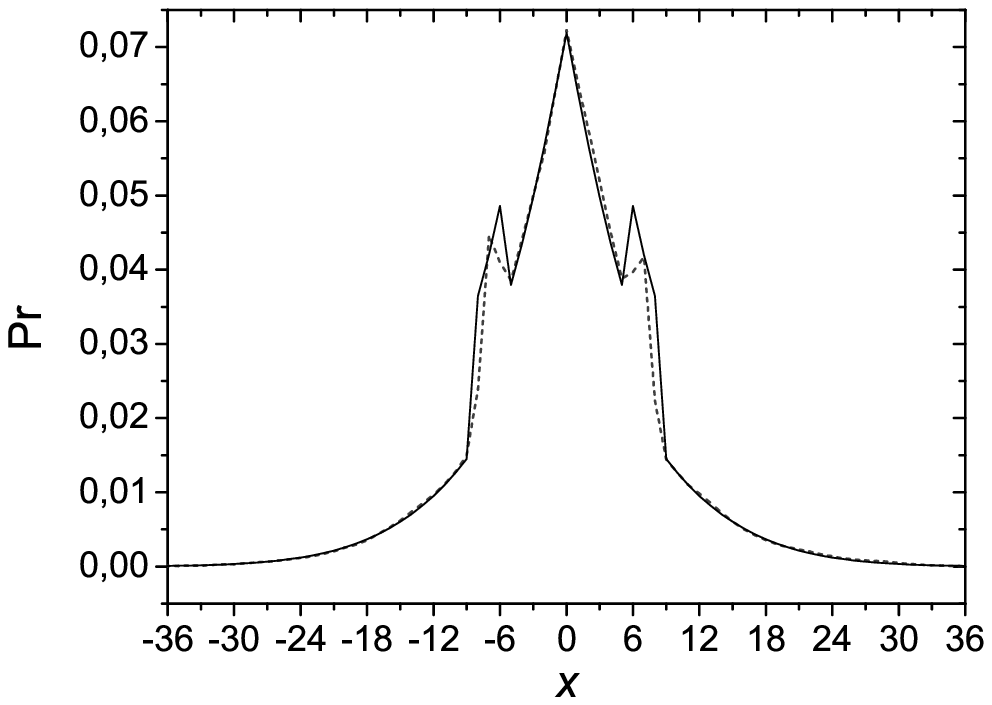}
\end{center}
\vspace*{-3mm}
\caption{\small Probability to find a walker at point $x\in{\mathbb Z}$ for $t=200$ (the total
number of steps) in three-zone environment. Solid line is obtained with the use of (\ref{dist}), while the dashed
line gives the result of averaging over 3000 random trajectories. Left panel: $2p_1=0.4$, $2p_2=0.9$, $2p_3=0.6$.
Right panel: $2p_1=0.6$, $2p_2=0.4$, $2p_3=0.9$. In the both cases, $a_1=6$, $a_2=8$.}
\end{figure}

Of course, the number of independent zones decreases if the neighbouring zones have the same
value of $p_n$ and, therefore, are joint. Interface point $x=a_n$ between zones with $p_{n}$
and $p_{n+1}=p_n$ becomes regular.

Note that (\ref{dist}) is similar to the amplitude squared in WKB approximation of quantum
mechanics, and the points $x=\{\pm a_n\}$ are connected with the turns, induced by a potential
in Schr\"odinger equation. In principal, a ``potential'' \cite{Gungor} in our model could appear by
excluding term $\partial_y{\cal P}$ in (\ref{DP}) and would be determined by Schwarzian
derivative of function (\ref{Vpot})
\begin{equation}\label{ffunc}
f(y)=\int_0^y{\tilde D}(1/2,y^\prime)\rmd y^\prime.
\end{equation}

In order to get expression for probability ${\rm Pr}(x,t)$, we should simply replace function
$\varphi(y,t)$ with $\Phi(y,t)$ in (\ref{dist}). To test our buildings, we compare the obtained
${\rm Pr}(x,t)$ with numerical simulations. Data are presented in figure~1. Analytical solution
at $\kappa=1$ is in agreement here with RW developing in three-zone environment. However, one
needs to account for higher-order corrections to reproduce better the considerable changes of
the probability profile at short $\Delta x$.

%%%%%%%%%%%%%%%%%%%%%%%%%%%%%%%%%%%%%%%%%%%%%%%%%%%%%%%%%%%%%%%%%%%%%%%%%%%%%%%%%%%%%%%%%%%%%
\section{Variance}

To investigate the walker diffusion, let us calculate the variance $\langle x^2\rangle$,
that is,
\begin{equation}
\Lambda(t)\equiv\int x^2\rmd\mu_t;\qquad
\langle x\rangle\equiv\int x\rmd\mu_t=0.
\end{equation}
Note that $\Lambda(t)$ is linear in $t$ in the case of purely Gaussian distribution.

In our case, one has $\Lambda(t)=\Lambda_0(t)+\kappa\Lambda_1(t)+O(\kappa)$, where
\begin{equation}
\Lambda_0(t)=\int_{-\infty}^{\infty}[x(y)]^2\varphi(y,t)\rmd y,\quad
\Lambda_1(t)=\int_{-\infty}^{\infty}[x(y)]^2{\cal S}_1(y,t)\rmd y.
\end{equation}

Although function $x(y)$ can be presented similarly to (\ref{Yfunc}), it is usefully
to re-write it in integrals in terms of $\{\tilde\chi_n(y)\}$. We find that
\begin{equation}
[x(y)]^2=y^2A_2(y)+|y|A_1(y)+A_0(y),\qquad
A_s(y)=\sum\limits_{n=1}^NA_{s,n}\tilde\chi_n(y);
\end{equation}
where numeric coefficients are
\begin{eqnarray}
&&A_{2,n}=p_n,\quad
A_{1,n}=2(\sqrt{p_n}c_n-p_nb_{n-1}),\quad
A_{0,n}=\left(c_n-\sqrt{p_n}b_{n-1}\right)^2,\\
&&c_n=\sum\limits_{m=1}^{n-1}\sqrt{p_m}(b_m-b_{m-1}).
\end{eqnarray}
Note that $A_2(y)$ coincides with diffusivity ${\tilde D}(y)$.

Combining the environment parameters $A_{s,n}$ and time dependent integrals, we write down
the variance components:
\begin{eqnarray}
\Lambda_0(t)&=&2\sum\limits_{s=0}^2\sum\limits_{n=1}^NA_{s,n}[U_s(b_n,t)-U_s(b_{n-1},t)],\\
\Lambda_1(t)&=&-2\sum\limits_{s=0}^2\sum\limits_{n=2}^NA_{s,n}B_n[U_s(b_n,t)-U_s(b_{n-1},t)]
\nonumber\\
&&+\sum\limits_{s=0}^2\sum\limits_{n=1}^N\sum\limits_{m=1}^{N-1} A_{s,n}\beta_m[V_s(b_n,t|b_m)-V_s(b_{n-1},t|b_m)],
\end{eqnarray}
where $B_n=\sum_{m=1}^{n-1}\beta_n$ for $n\geq2$, and
\begin{eqnarray}
U_s(b,t)&\equiv&\int_0^by^s\varphi(y,t)\rmd y,\\
V_s(B,t|b)&\equiv&\int_{0}^{B}y^s[\varphi(|y-b|+b,t)+\varphi(|y+b|+b,t)]\rmd y.
\end{eqnarray}

The Gaussian integrals yield
\begin{eqnarray}
&U_0(b,t)=\frac{1}{2}{\rm erf}\left(\frac{b}{2\sqrt{t}}\right),\quad
U_1(b,t)=\sqrt{\frac{t}{\pi}}\left[1-\exp{\left(-\frac{b^2}{4t}\right)}\right],&
\nonumber\\
&U_2(b,t)=t~{\rm erf}\left(\frac{b}{2\sqrt{t}}\right)
-b\sqrt{\frac{t}{\pi}}\exp{\left(-\frac{b^2}{4t}\right)}.&
\end{eqnarray}

Using $U_s(b,t)$, we also get
\begin{eqnarray}
V_0(B,t|b)&=&U_0(B+2b,t)-\theta(b-B)U_0(2b-B,t)
\nonumber\\
&&+\theta(B-b)\left[U_0(B,t)-2U_0(b,t)\right],\\
V_1(B,t|b)&=&U_1(B+2b,t)-2U_1(2b,t)+4bU_0(2b,t)-2bU_0(B+2b,t)
\nonumber\\
&&+\theta(b-B)\left[U_1(2b-B,t)-2bU_0(2b-B,t)+2bU_0(b,t)\right]
\nonumber\\
&&+\theta(B-b)U_1(B,t),\\
V_2(B,t|b)&=&U_2(B+2b,t)-2U_1(2b,t)-4bU_1(B+2b,t)
\nonumber\\
&&+4b^2U_0(B+2b,t)
\nonumber\\
&&-\theta(b-B)\left[U_2(2b-B,t)-4bU_1(2b-B,t)\right.
\nonumber\\
&&\left.+4b^2U_0(2b-B,t)\right]
\nonumber\\
&&+\theta(B-b)\left[U_2(B,t)-2U_2(b,t)+4bU_1(b,t)\right.
\nonumber\\
&&\left.-4b^2U_0(b,t)\right],
\end{eqnarray}
where $\theta(0)=1/2$ is assumed.

\begin{figure}
\begin{center}
\includegraphics[width=7.55cm]{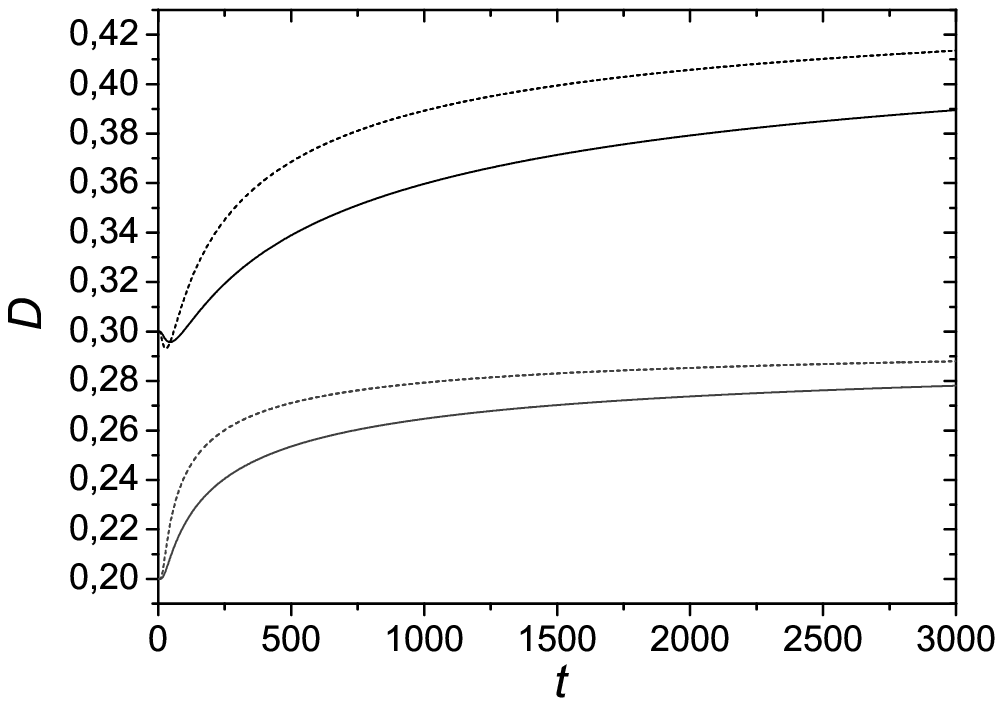}
\includegraphics[width=7.3cm]{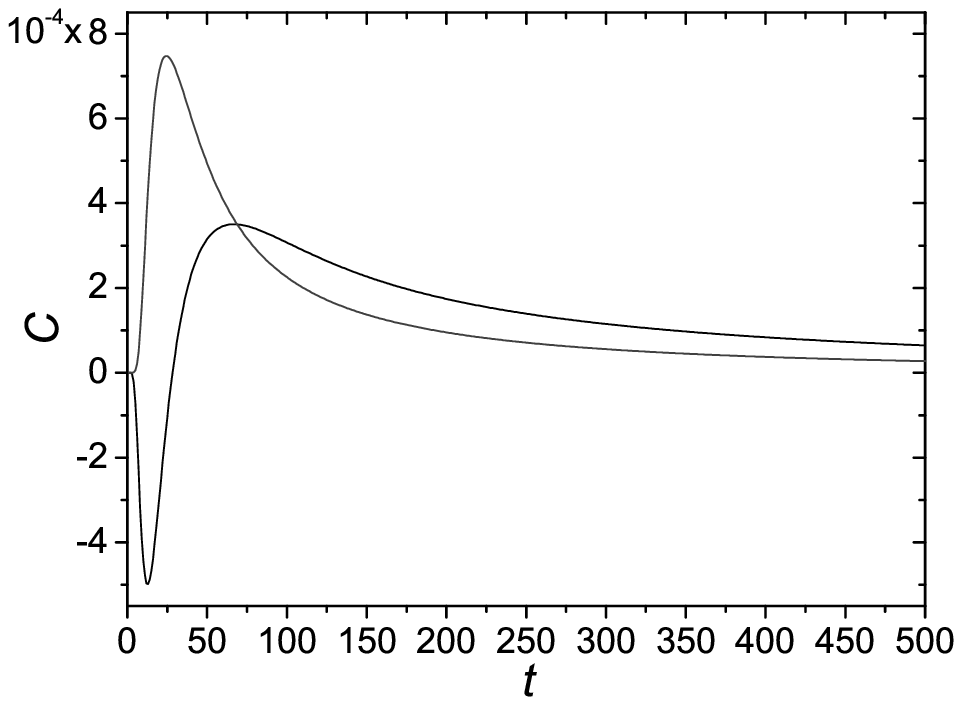}
\end{center}
\vspace*{-3mm}
\caption{\small Left panel: Time dependence of effective diffusivities $\Lambda/2t$
(solid line) and $\partial_t\Lambda/2$ (dashed line) for two sets of parameters as in figure~1.
Right panel: Dependence of $C\equiv\partial^2_t\Lambda/2$ on $t$. Gray curves are for model
with $p_1=0.2$, while black ones are for $p_1=0.3$.}
\end{figure}

Analyzing $\Lambda(t)$, let us introduce effective diffusivities $\Lambda(t)/2t$ and
$\partial_t\Lambda(t)/2$. Thus, if $p_n=p$ for all $n$, these coincide with $p$.

We also consider the second-order derivative of $\Lambda(t)$, denoted by $C$ and connected with
velocity autocorrelation function accordingly to Kubo's relation~\cite{Kubo66,BTN}:
\begin{equation}
\frac{1}{2}\frac{\rmd\langle x^2(t)\rangle}{\rmd t}=
\int_0^t\langle\dot x(t^\prime)\dot x(0)\rangle\rmd t^\prime,\qquad
t\to\infty,
\end{equation}
where $x(t)$ is the continuous stochastic coordinate with $\langle x(t)\rangle=0$
and $\langle \dot x(t)\rangle=0$. 

The results based on our formulae are sketched in figure~2. We see on the left panel that there are
the time intervals where the effective diffusivities are decreasing and increasing. These may correspond,
respectively, to transient regimes of subdiffusion and superdiffusion, represented by $C<0$ and $C>0$ on
the right panel. We conclude that a role of $n$-th zone in the diffusion picture depends mainly
on adsorption effect which is proportional to probability $r_n=1-2p_n$. Moreover, the model situations
under consideration demonstrate inaccessibility of the regime with $p_3$ in a given time interval.
Although, putting $b_N\to\infty$ before tending $t$ to infinity, it is expected that
the only terms with $U_2(b_N,t)/t\to1$ should survive at $t\to\infty$ in a strong mathematical
sense. However, we found that this limit is slowly reached because of environment structure.
In other words, there is always a non-vanishing probability of walker capture by $n$-th zone
with $r_n>r_N$. This determines the time delay which may be estimated. Moreover, we assume an
existence of different time scales which should be accurately defined~\cite{DLKB} for a deeper
understanding.

On the other hand, the right panel of figure~2 illustrates that the correlation $C$ is relaxed with time,
and information about early stage of evolution, developed in region near the origin, is lost. Being in
infinitely wide $N$-th zone for a long time, a role of whole zone structure decreases in a walker
history.

It is interesting to note that stage with $C<0$ is often connected with a presence of liquid-like state
in many-particle system~\cite{BTN}. This state emerges here due to specifically located zone with $2p=0.4$
and could be intuitively foreseen, although sufficient conditions of its appearance are still unclear for us.
We can only say that the zones with relatively small $p_n$ keep a walker in a trap for a short time,
so after a while the walker is able to diffuse away.

%%%%%%%%%%%%%%%%%%%%%%%%%%%%%%%%%%%%%%%%%%%%%%%%%%%%%%%%%%%%%%%%%%%%%%%%%%%%%%%%%%%%%%%%%%%%%
\section{Discussion}

Initially, we have formulated the model of one-dimensional RW with space dependent transition probability which is
identical for the left and right steps. We describe a wandering in space with the finite-sized zones having
the properties of adsorption in the bulk and partial reflection at the boundary points.
It generalizes a problem of RW with various barriers \cite{Gupta,Percus,Ascher60} and allows us to speak
about heterogeneous environment. Locating the  $N$ zones $x\in(-a_n,-a_{n-1})\cup(a_{n-1},a_n)$ densely along
coordinate axis, transition probability in $n$-th zone is determined by constant parameter $p_n$, except
the separation points where its value changes. This assumption is crucial for finding an analytical
solution to the model equations in diffusion approximation at the large number of walker steps.
Step-like dependence of diffusion coefficient on $x$ prompts us to introduce the new coordinate $y(x)$,
in terms of which the probability distribution is constructed with the use of Gaussian one (\ref{dist}).
Nevertheless, the obtained distribution function is non-Gaussian because of complexity of environment
and surface effects at singular points $\{\pm a_n\}$ where the derivatives of functions with respect
to $x$ diverge.

It seems to be difficult to describe analytically an effect of infinitely thin zone (a wall) without additional suggestions.
Similar problem arises in zone interface description. Comparing the numerical RW simulations with discrete
$x\in{\mathbb Z}$ and analytical solutions in continuous limit, we might effectively displace a position $a_n$
of point-like barrier up to $a_{n}\pm1/2$ in order to obtain the finitely wide border. Another possibility
is to use the solutions with a gap \cite{Powels} which are not considered here.

Actually, our model describes the diffusion process, taking into account the long-range properties of environment
and neglecting local fluctuations $\delta D(x)\equiv D(x)-{\bar D}(x)$ of smooth-varied diffusivity $D(x)$, where
\begin{equation}
\bar D(x)=\sum\limits_{n=1}^N D_n\chi_n(x),\qquad
D_n=\frac{1}{2(a_n-a_{n-1})}\int_{-\infty}^\infty D(x)\chi_n(x)\rmd x.
\end{equation}

Analyzing the results presented in figure~1, we can clarify the physical reasons of non-trivial distribution
of the particles emanating from the origin.

Explanation of remarkable probability changes is based on a role of walker adhesion, which is
forced by environment and related with probability $r_n=1-2p_n$. Sequence of zones with different
values of $p_n$ is also significant: the sign of difference $p_{n+1}-p_n$ is responsible for preferable
direction of the walk on interface of neighboring zones indexed by $n$ and $n+1$. Relative time of
the walker being in $n$-th zone depends on the number $n$, width $a_n-a_{n-1}$, and $r_n$. Thus, adhesion
effect can be evaluated by characteristic time and length. Nevertheless, a long-time tail of
the variance corresponds to the ordinary diffusion with $\langle x^2\rangle\sim t$ at $t\to\infty$.

Alternative treatment can be done in terms of mechanics by introducing an effective walker ``mass'' $m_n=1/2p_n$
for each zone of environment. Considering a macroscopic ensemble of the particles, we can interpret the mass
changes as the result of geometrically dependent interaction among particles which is not specified in our
picture but leads to appearance of the different states or phases confined within the given intervals.

It is expected that the Fourier transform ${\tilde P}(k,t)$ of PDF can give us the one-particle distribution
function in momentum space $k$, accounting for the phase transitions in our case, in accordance with ideas
of Bloch and Balescu. Indeed, replacing $t$ and $p_n$ with inverse temperature $1/T$ and $1/2m_n$,
respectively, PDF of ordinary diffusion process,
\begin{equation}
{\tilde P}(k,t)\propto\rme^{-k^2pt},
\end{equation}
plays a role of Boltzmann distribution, which is a starting point of deriving thermodynamic functions
and relations for many-particle ensemble.

%%%%%%%%%%%%%%%%%%%%%%%%%%%%%%%%%%%%%%%%%%%%%%%%%%%%%%%%%%%%%%%%%%%%%%%%%%%%%%%%%%%%%%%%%%%%%
\ack

A.N. thanks to A.M.~Gavrilik (BITP) for fruitful discussions and to Yu.M.~Sinyukov (BITP) for
suggestion to highlight some aspects. A.N. is indebted for partial support of this work
by Department of Physics and Astronomy of NAS of Ukraine.

%%%%%%%%%%%%%%%%%%%%%%%%%%%%%%%%%%%%%%%%%%%%%%%%%%%%%%%%%%%%%%%%%%%%%%%%%%%%%%%%%%%%%%%%%%%%%
\appendix
\section{Potential Model}

Let us substitute ${\cal P}(y,t)=\exp{[-B(y)/2]}\Pi(y,t)$ in (\ref{DP}) at
$\kappa=1$. One gets
\begin{equation}
-\partial_t\Pi=-\partial^2_y\Pi+V\Pi.
\end{equation}
Potential $V$ can be presented as
\begin{equation}\label{Vpot}
V(y)\equiv-\frac{1}{2}\left(\beta^\prime-\frac{1}{2}(\beta)^2\right)=-\frac{1}{2}\mathrm{S}f(y),
\end{equation}
where $\beta(y)$ and $f(y)$ are given by (\ref{beta}) and (\ref{ffunc}), respectively.
We use formally notation for Schwarzian derivative 
$\mathrm{S}f\equiv(f^{\prime\prime}/f^{\prime})^{\prime}-(f^{\prime\prime}/f^{\prime})^2/2$;
and relation $\beta=f^{\prime\prime}/f^{\prime}$.

Note that the form of (\ref{Vpot}) is similar to the potential of super-symmetric problem~\cite{BP}.
Function $u(y)\equiv1/\sqrt{f^\prime}={\tilde D}(-1/4,y)$ is a solution to stationary equation
\begin{equation}
u^{\prime\prime}+\frac{1}{2}(\mathrm{S}f)u=0.
\end{equation}

%%%%%%%%%%%%%%%%%%%%%%%%%%%%%%%%%%%%%%%%%%%%%%%%%%%%%%%%%%%%%%%%%%%%%%%%%%%%%%%%%%%%%%%%%%%%%

\end{document}